\UseRawInputEncoding
\documentclass{webofc}
\usepackage[varg]{txfonts}   
\usepackage{subcaption}
\usepackage{graphicx}

\newcommand\Tstrut{\rule{0pt}{2.6ex}}           
\newcommand\Bstrut{\rule[-0.9ex]{0pt}{0pt}}     

\begin{document}
\title{Short-lived Radionuclides in the Milky Way Galaxy}
%
%

\author{\firstname{Tejpreet} \lastname{Kaur}\inst{1}\fnsep\thanks{\email{tejpreet@pu.ac.in;tejpreetkaur95@gmail.com}} 
}

\institute{Department of Physics, Panjab University, Chandigarh, 160014, India 
          }

\abstract{%
The short-lived radionuclides (SLRs) have a half-life $\leq$ 100 Myr. The $\gamma$-ray observations and excess abundance of their daughter nuclides in various meteoritic phases confirm the existence of SLRs in the Galaxy and early solar system (ESS), respectively. In this work, we have developed Galactic Chemical Evolution (GCE) models for SLRs, $^{26}$Al, and $^{60}$Fe along with $^{36}$Cl, $^{41}$Ca, and $^{53}$Mn. These models predict the temporal and spatial evolution of SLR abundance trends in the Galaxy from 2-18 kpc. The abundance of two SLRs, $^{26}$Al, and $^{60}$Fe, are investigated further, as their $\gamma$-ray observations are available for comparison with the model predictions. The predictions for the abundance per unit area for each ring decrease from the inner to outer regions of the Galaxy.The GCE predictions for the total mass of alive $^{26}$Al, and $^{60}$Fe in 2-18 kpc of the Galaxy at present time are 0.2 M$_\odot$ and 0.08 M$_\odot$, respectively.

}
\maketitle
\section{Introduction}
Short-lived radionuclides (SLRs) are elements with a half-life of the order of a few million years. These nuclides can be formed in various stellar environments in the Galaxy. The presence of the SLRs is vital evidence that star formation is an ongoing process in the Galaxy. The main production sites of $^{26}$Al are Asymptotic Giant Branch stars (AGB), Core collapse Supernovae (CCSNe), nova and Wolf-Rayet (WR) stars. For $^{60}$Fe, the primary production happens during the neutron capture reaction inside the CCSNe. These are the only two short-lived radionuclides for which the $\gamma$-ray observations can provide evidence of their presence in the star--forming regions. For the other SLRs, $^{36}$Cl, $^{41}$Ca and, $^{53}$Mn, CCSNe are major production sites except for $^{53}$Mn, which can also be produced in Supernova Ia (SNIa). The details of the characteristics of these nuclides are given in Table \ref{table:1}. The $\gamma$-ray COMPTEL and INTEGRAL observations detected the presence of $^{26}$Al in the Galaxy. The results show the map of $^{26}$Al which is concentrated in the galactic plane \citep{comptel26al2001ESASP.459...55P} \citep{integralDiehl_2013}. The $^{26}$Al-- rich regions in the $\gamma$-ray map coincide with the star-forming regions of the disc region. The $^{26}$Al emits $\gamma$-rays when it decays to $^{26}$Mg at 1809 keV. The $^{26}$Al emission regions also contain the observations of the presence of $^{60}$Fe, which emits $\gamma$-rays while decaying into $^{60}$Co and further $^{60}$Ni at 1173 keV and 1332 keV, respectively. With the ability of INTEGRAL to resolve the star-forming region, the $^{26}$Al can be observed in the stellar ejecta, and nucleosynthesis studies can be performed inside the star-forming regions. The observations of $^{26}$Al and $^{60}$Fe from the same star-forming regions are being used to constrain the stellar nucleosynthesis models, identifying more such regions will also help identify the source of these SLRs in the ESS.  Also, the observations from the other galaxies will provide more stringent constraints for the stellar and galactic chemical evolution modelling community. We have studied the abundance of SLRs using the Galactic Chemical Evolution (GCE) models discussed in \citep{Kaur2019MNRAS.490.1620K}.


\begin{table}[h!]
\centering
\begin{tabular}{||c| c| c| c| c| c||} 
 \hline
 Short-lived radionuclide (SLR) & $^{26}$Al & $^{36}$Cl & $^{41}$Ca & $^{53}$Mn & $^{60}$Fe \Tstrut \\ [0.5ex] 
 \hline\hline
 Half-life (t$_{1/2}$
(Myr)) & 0.717 & 0.301 & 0.0994 & 3.74 & 2.62 \Tstrut\Bstrut \\ 
 \hline
 Mean-life($\tau $ (Myr)) & 1.035 & 0.434 & 0.1434 & 5.40 &  3.78 \Tstrut\Bstrut \\ 
 \hline
 Decay Process & $\beta$$_{+}$ & $\beta$$_{-}$ & $\beta$$_{+}$ & $\beta$$_{+}$  & $\beta$$_{-}$ \Tstrut\Bstrut \\ 
 \hline
 Daughter Product & $^{26}$Mg & $^{36}$S, $^{36}$Ar & $^{41}$K & $^{53}$Cr & $^{60}$Ni \Tstrut\Bstrut \\ 
 \hline
 Stable Isotope & $^{27}$Al & $^{35}$Cl & $^{40}$Ca & $^{55}$Mn & $^{56}$Fe \Tstrut\Bstrut \\  
 \hline
 
\end{tabular}
\caption{Details of Short-Lived Radio nuclides considered in this work. The data of half-life and decay modes is taken from \citep{LUGARO20181}.}
\label{table:1}
\end{table}

\section{Galactic Chemical Evolution Models}
The study of the abundance distribution of SLRs plays a vital role in understanding the  Galaxy's active star-forming regions. The SLR prediction for the Solar System also gives insights into its origin and the pre-solar molecular cloud. Despite precise observations of SLR abundances in the Solar System from meteorites, their stellar sources are still debated. To explain the Galactic scale steady-state distribution of SLRs, we divided the Galaxy into rings of 2 kpc width from 2–18 kpc, based on Monte Carlo simulations described in \citep{SK18} \& \citep{kaur_2021}. The homogeneous GCE model presented here is based on three-infall accretion for the formation of the Galaxy. In the case of the three infall model, the halo, thick and the thin disc all form in separate accretion episodes. The first two episodes occur on a shorter time scale, followed by the third episode in which thin disc forms over the extended time scale in inside--out scenario \cite{Micali2013} \cite{chiappini1997}. In inside-out scenario the inner regions of the thin disc forms first and then the outer regions. The normalisation constants involved in the accretion are determined by reproducing the observed value of total surface mass density in the solar neighbourhood and other parts of the Galaxy. Various generations of stars are formed according to the prevailing star formation rate (SFR), a function of the gas and prevalent total surface mass densities. Also, the temporal and radial star formation efficiencies, $\nu (t)$ and $\eta(r)$, respectively, account for the temporal and spatial variation in the SFR. The stellar mass in each ring at every time step is distributed in the stars according to the initial mass function (IMF) in the mass range 0.8–100 M$_\odot$ in various stellar generations. These stars evolve according to their mass and metallicity. Then, at every time step T, an assessment is made for all stars formed before T (t$<$T), to estimate their nucleosynthetic yields and remnants.  A binary stellar population is synthesised at the time of star formation. With the binary fraction f, some stars out of the total stars formed at each time step evolve into a definite progenitor of SNe Ia. The progenitor stars have mass in the range of 3–8 M$_\odot$ and 11–16 M$_\odot$. The radioactive nuclides produced are subject to decay at each time step. The stellar yields of AGB and massive stars contribute the radioactive nuclides along with the stable nuclides to the interstellar medium (ISM). The contribution to the ISM for CCSNe is considered from \citep{Woosley1995}, and AGBs from \citep{karakas2010} and \citep{Cristallo_2015}. 
 
\label{model}
\begin{figure}[!tbp]
  \centering
  \subfloat{\includegraphics[width=0.45\textwidth]{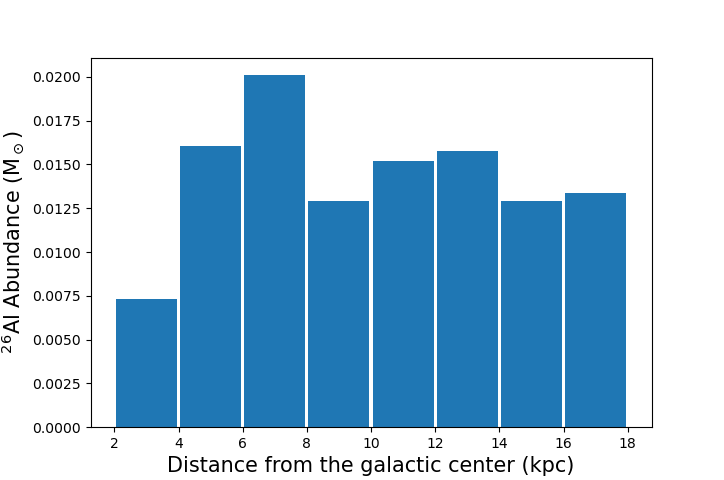}} 
  \hfill
  \subfloat{\includegraphics[width=0.45\textwidth]{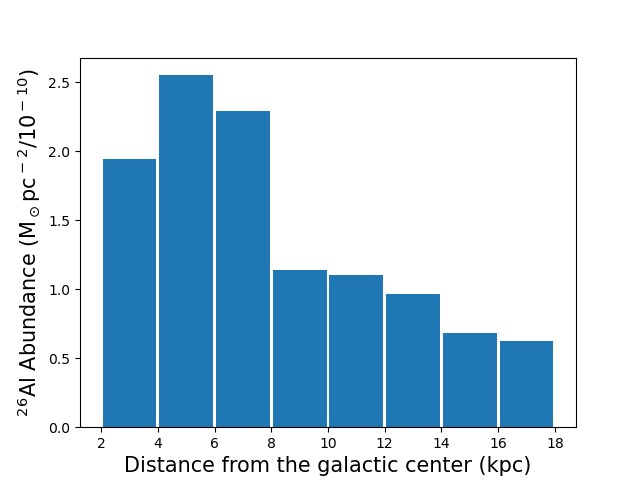}} 
  \caption{(Left) Absolute abundance of $^{26}$Al abundance, (Right)  $^{26}$Al per unit area in the Galaxy for eight annular rings from 2-18 kpc. }
\end{figure}

\begin{figure}[!tbp]
  \centering
  \subfloat{\includegraphics[width=0.45\textwidth]{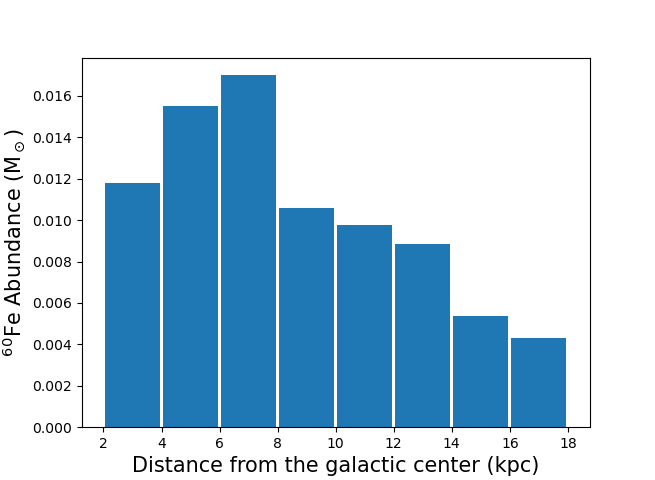}} 
  \hfill
  \subfloat{\includegraphics[width=0.45\textwidth]{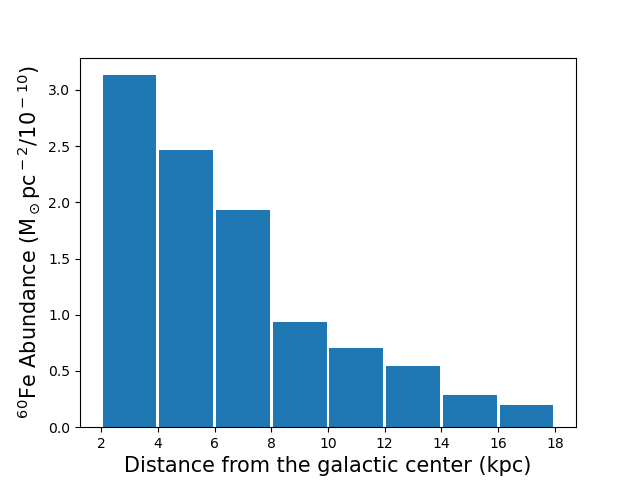}}
  \caption{(Left) Absolute abundance of $^{60}$Fe abundance, (Right)  $^{60}$Fe per unit area in the Galaxy for eight annular rings from 2-18 kpc.}
\end{figure}

\section{Results and Discussion}

Results shown in figures 1-3 are based on the models I of homogeneous GCE models presented in \citep{Kaur2019MNRAS.490.1620K}. Figure 1 shows the absolute abundance and abundance per unit area for $^{26}$Al in the Galaxy. Similar results for $^{60}$Fe are presented in Figure 2. Then the ratio of the absolute abundance of $^{60}$Fe and $^{26}$Al in the Galaxy is presented in Figure 3.  The GCE model shown here evolves each ring of the Galaxy of width 2 kpc from 2-18 kpc over the galactic time scale. The model also predicts the metallicity and [Fe/H] for the Galaxy over the galactic time scale. Each ring experiences the accretion of gas and forms various generations of stars as explained in section \ref{model} above and in \citep{Kaur2019MNRAS.490.1620K}. The SLRs trends mainly depend upon the star formation rate in the region,which is higher in the inner regions and decreases as we move towards the outer regions of the Galaxy. This trend is mainly because of the higher accretion in the inner regions and high the star formation efficiency. However, $^{26}$Al does not follow the decreasing trend as well as $^{60}$Fe because of the contribution of $^{26}$Al from AGB stars. On the contrary, $^{60}$Fe is mainly coming from massive stars. The sum of the mass of $^{26}$Al and $^{60}$Fe over the 2-18 kpc of the Galaxy is ~0.2 M$_\odot$ and 0.08 M$_\odot$, respectively.

The abundance of other SLRs, $^{36}$Cl, $^{41}$Ca and $^{53}$Mn, also follow a similar trend of decrease in the abundance per unit area of the ring as we move towards the Galaxy's outer regions as shown in figure 1 of \citep{Kaur2019MNRAS.490.1620K}. The areas of the rings from 2-18 kpc, which have 2 kpc width, from the inner first to outer eighth ring, are given as 37.68, 62.8, 87.92, 113.04, 138.16, 163.28, 188.4, and 213.52 kpc$^2$ respectively. In this GCE model, the inventories of every generation of stars are all mixed with the ISM gas and homogenised with the entire ring area. Hence abundance per unit area gives better insight into the abundance distribution of the SLRs in the Galaxy. 

The $^{26}$Al and $^{60}$Fe can be synthesised inside two different evolutionary phases of the massive stars \cite{DIEHL2022103983}. So there are suggestions that the ratio of these two SLRs $^{26}$Al and $^{60}$Fe can be an important indicator of the stellar population in the star-forming region. The $\gamma$ ray observations show that the flux ratio of $^{60}$Fe/$^{26}$Al is 0.4 , which limits the steady--state mass ratio for $^{60}$Fe/$^{26}$Al to be $<$ 0.9 \cite{DIEHL2022103983}. Figure 3 shows the trend of this ratio with the distance from the galactic center. There is significant decrease in the ratio with an increase in distance from the galactic center. The ratio is $<$ 0.9 in the Galaxy except for the first two galactic rings from 2-6 kpc. In the solar neighborhood the ratio is $\approx$ 0.8 and decreases afterwards in the outer regions which could arise from the AGB contribution to $^{26}$Al abundance in the Galaxy. 

\label{sec-1}

\begin{figure}[]
\includegraphics[width=0.5\textwidth]{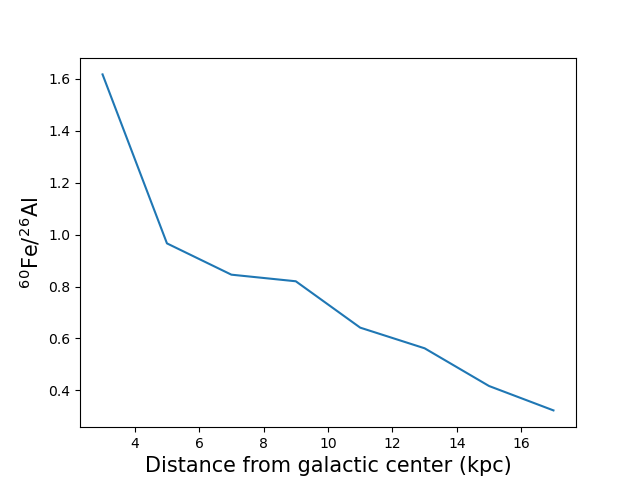}
\centering
\caption{The ratio of absolute abundance of $^{60}$Fe/$^{26}$Al in the Galaxy for eight annular rings from 2-18 kpc.}
\label{Fealratio}       
\end{figure}

\section{Conclusion}
The SLRs abundance distribution presented here are based on the homogeneous GCE models. The GCE models represent a good approximation to understand the overall trend in the Galaxy \cite{Kaur2022MmSAI..93d..29K}. The homogeneous models show the abundance per unit area of $^{26}$Al and $^{60}$Fe in the Galaxy is higher in the the inner regions and lower in the outer. This trend is in agreement with the $\gamma$-ray observation. The estimates for the mass of $^{26}$Al and $^{60}$Fe in the entire Galaxy from the $\gamma$-ray observations are 1.8–2.4 M$_\odot$, and  1–6 M$_\odot$, respectively, which are higher than the predicted values from the GCE model explained above \cite{thesisal}. The low mass of SLRs could be outcome of the approach used to mix the stellar ejecta with the ISM. The stellar ejecta of every star is mixed with the entire  annulus ring taken into account. This approximation is valid for the stable nuclides and have a little effect on their abundance predictions. However, in case of the SLRs, because of the short half-life the mixing of the ejecta with ISM can significantly reduce the abundance predictions. The second reason for low predictions from the GCE model is that the central 2 kpc is excluded from the model calculation. Finally, the ratio of $^{60}$Fe/$^{26}$Al from the GCE can provide clues to the contribution of different stellar sources to these two SLRs.

\section{Acknowledgement}
TK is thankful to Dr Thomas Siegert from the University of Würzburg, Germany, for valueable discussions and suggestions related to the $\gamma$-ray observations for SLRs in the Milky Way Galaxy. This manuscript was written during a visit to the Konkoly Observatory in Budapest, Hungary, hosted by Dr Maria Lugaro, for which TK is also grateful.

\bibliography{NPA_X}

\begin{thebibliography}{14}

\bibitem{comptel26al2001ESASP.459...55P}
S.~{Pl{\"u}schke}, R.~{Diehl}, V.~{Sch{\"o}nfelder}, H.~{Bloemen},
  W.~{Hermsen}, K.~{Bennett}, C.~{Winkler}, M.~{McConnell}, J.~{Ryan},
  U.~{Oberlack} et~al., \emph{{The COMPTEL 1.809 MeV survey}}, Vol. 459 of
  \emph{ESA Special Publication} (2001)

\bibitem{integralDiehl_2013}
R.~Diehl, Reports on Progress in Physics \textbf{76}, 026301 (2013)

\bibitem{Kaur2019MNRAS.490.1620K}
T.~{Kaur}, S.~{Sahijpal}, Monthly Notices of the Royal Astronomical Society
  \textbf{490}, 1620 (2019)

\bibitem{LUGARO20181}
M.~Lugaro, U.~Ott, {\'A}.~Kereszturi, Progress in Particle and Nuclear Physics
  \textbf{102}, 1 (2018)

\bibitem{SK18}
S.~Sahijpal, T.~Kaur, Monthly Notices of the Royal Astronomical Society
  \textbf{481}, 5350 (2018)

\bibitem{kaur_2021}
T.~Kaur, Ph.D. thesis, Panjab University (2021),
  \urlstyle{tt}\url{http://hdl.handle.net/10603/331671}

\bibitem{Micali2013}
A.~{Micali}, F.~{Matteucci}, D.~{Romano}, Monthly Notices of the Royal
  Astronomical Society \textbf{436}, 1648 (2013)

\bibitem{chiappini1997}
C.~Chiappini, R.~Gratton, The Astrophysical Journal \textbf{477}, 765 (1997)

\bibitem{Woosley1995}
S.E. {Woosley}, T.A. {Weaver}, Astrophysical Journal Supplement \textbf{101},
  181 (1995)

\bibitem{karakas2010}
A.I. {Karakas}, Monthly Notices of the Royal Astronomical Society \textbf{403},
  1413 (2010)

\bibitem{Cristallo_2015}
S.~Cristallo, O.~Straniero, L.~Piersanti, D.~Gobrecht, The Astrophysical
  Journal Supplement Series \textbf{219}, 40 (2015)

\bibitem{DIEHL2022103983}
R.~Diehl, A.J. Korn, B.~Leibundgut, M.~Lugaro, A.~Wallner, Progress in Particle
  and Nuclear Physics \textbf{127}, 103983 (2022)

\bibitem{Kaur2022MmSAI..93d..29K}
T.~{Kaur}, D.~{Romano}, M.~{Molero}, A.~{Vasini}, F.~{Matteucci}, Memorie della
  Societa Astronomica Italiana \textbf{93}, 29 (2022)

\bibitem{thesisal}
M.M.M. Pleintinger, Ph.D. thesis, Technische Universität München, München
  (2020)

\end{thebibliography}

\end{document}